\documentclass[conference]{IEEEtran}
\IEEEoverridecommandlockouts
\usepackage{cite}
\usepackage{physics}
\usepackage{amsmath,amssymb,amsfonts}
\usepackage{algorithm}
\usepackage[noend]{algpseudocode}
\usepackage{graphicx}
\usepackage{textcomp}
\usepackage{xcolor}
 \usepackage{float} 
\def\BibTeX{{\rm B\kern-.05em{\sc i\kern-.025em b}\kern-.08em
    T\kern-.1667em\lower.7ex\hbox{E}\kern-.125emX}}
\DeclareMathOperator*{\argmin}{arg\,min}
\DeclareMathOperator*{\mintime}{min}
\begin{document}

\title{How viable is quantum annealing for solving linear algebra problems?}

\author{\IEEEauthorblockN{Ajinkya Borle}
\IEEEauthorblockA{\textit{CSEE Department} \\
\textit{University of Maryland, Baltimore County}\\
Baltimore, USA \\
aborle1@umbc.edu}
\and
\IEEEauthorblockN{Samuel J. Lomonaco}
\IEEEauthorblockA{\textit{CSEE Department} \\
\textit{University of Maryland, Baltimore County}\\
Baltimore, USA \\
lomonaco@umbc.edu}
}

\maketitle

\begin{abstract}
With the increasing popularity of quantum computing and in particular quantum annealing, there has been growing research to evaluate the meta-heuristic for various problems in linear algebra: from linear least squares to matrix and tensor factorization. At the core of this effort is to evaluate quantum annealing for solving linear least squares and linear systems of equations. In this work, we focus on the viability of using quantum annealing for solving these problems. We use simulations based on the adiabatic principle to provide new insights for previously observed phenomena with the D-wave machines, such as quantum annealing being robust against ill-conditioned systems of equations and scaling quite well against the number of rows in a system. We then propose a hybrid approach which uses a quantum annealer to provide a initial guess of the solution $x_0$, which would then be iteratively improved with classical fixed point iteration methods.
\end{abstract}


\section{Introduction}
Linear Algebra as a branch of mathematics is ubiquitous in its applications in many fields \cite{boyd2018introduction}. As an example, it helps us represent, manipulate and make predictions on data, just to name a few uses in machine learning and data science \cite{brownlee2018basics}. With the advent of quantum computing and the increasing availability of quantum computers, there has been considerable research done for applying it to linear algebra problems \cite{harrow2009quantum,wiebe2012quantum,bravo2019variational,clader2013preconditioned,chang2019quantum}. Most prominent among them being the Harrow-Hassidim-Llyod algorithm \cite{harrow2009quantum} for solving a system of linear equations, notable for its exponential speedup over classical methods.

However, this algorithm and other similar ones have a lot of assumptions that limit its near-term applicability at the very least\footnote{mentioned in the background section} \cite{aaronson2015read}. As an alternative, the discrete optimization meta-heuristic of quantum annealing is a method that stands out for being practical in the near-term future. At the time of writing this paper, we have seen multiple works relating to the formulation of linear algebra problems for quantum annealing and the use of D-wave quantum annealers for evaluating the results.

Since quantum annealing (QA) is a meta-heuristic, it becomes more challenging to ascertain if it is a viable approach for solving linear algebra problems, especially when problems like linear least squares (LLS) and linear systems of equations (LSE) can be solved in polynomial time classically. In this work, we want to share some insights on the viability of using QA for solving linear algebra problems.

We first conduct simulations using the adiabatic quantum computing (AQC) principle for LLS and see how the running-time gets affected by different factors such as number of variables, number of equations, precision of variables and the condition number. Our results show quantum annealing performing well for ill-conditioned systems and large number of rows, but it performs poorly if the precision of a variable is increased (exponential slowdown). This supports a few empirical results in previous works that briefly comment on similar phenomenon  observed on the D-wave quantum annealers \cite{chang2019quantum,chang2019least,potok2021adiabatic}. In our work, by analyzing the phenomenon at the level of principle (AQC simulation), we are able to make more accurate observations that are machine agnostic.

We then posit a hybrid approach for general LSE and LLS problems where we use D-wave quantum annealers to get a initial approximation $x_0$ for the unknown vector $x$ and then use classical fixed point techniques to solve the problems to the necessary error threshold. Our results seem to indicate that this may work better for LLS than LSE.

The rest of the report is organized as follows: Section \ref{sec:bgrw} deals with the background required for this paper as well as relevant work in this research direction. Section \ref{sec:aqc_sim} is on our work on adiabatic simulations to assess the growth of running-time with respect to different parameters of the LLS problem. Section \ref{sec:qals} is where we propose a hybrid quantum-classical approach, the quantum-assisted linear solver. We conclude the paper in Section \ref{sec:conc} with our final thoughts on the results of the paper.
\section{Background and related work}\label{sec:bgrw}
\subsection{Background}\label{sec:background}
In this part we'll introduce and briefly cover the terms and concepts involved in the work.
\subsubsection{Ising model}
This is a mathematical model that has its origins in statistical ferromagnetism \cite{gallavotti1999statistical}. Originally used to represent magnetic dipoles moments of atomic spins, in essence, it consists of discrete bipolar elements that can take a value of either -1 or +1. The Ising objective function is as follows:
\begin{align}
    F(\sigma) = \sum_{a}h_a\sigma_a + \sum_{a<b}J_{ab}\sigma_{a}\sigma_{b}\label{eq:ising}
\end{align}
where $\sigma_{a}\in \{-1,1\}$ is a bipolar variable ,  $h_a$ and $J_{ab}$  are the coefficients for the linear and quadratic terms respectively \cite{dorband2016stochastic} (or the external fields and interaction strengths in ferromagnetic terms).

Most important for our purposes, is that the Ising model's decision problem is NP-Complete in nature \cite{barahona1982computational}. The objective function has been used to model many NP-Hard problems as either minimization or maximization of Eqn(\ref{eq:ising}) (primarily the former) \cite{lucas2014ising}.

\subsubsection{Quadratic unconstrained binary optimization (QUBO)}

A quadratic unconstrained binary optimization (QUBO) is a combinatorial optimization problem equivalent to the Ising objective function Eqn(\ref{eq:ising}), but on binary variables with the following cost function:
\begin{align}
    F'(q) = \sum_{a}v_{a}q_{a} + \sum_{a<b}w_{ab}q_{a}q_{b}\label{eq:qubo}
\end{align}
here $q_{a} \in \{0,1\}$ are the binary variables returned by the machine after the minimization and $v_{a}$ and $w_{ab}$ are the coefficients for the qubits and the couplers respectively \cite{dorband2016stochastic}. 
The following relationship between $\sigma_{a}$ and $q_{a}$ shows the equivalence:
\begin{align}
    \sigma_{a} &= 2q_{a} -1 \label{eq:ising2qubo}\\
    \text{and } F(\sigma) &= F'(q) + \text{offset}
\end{align}
The offset value is a constant and the actual minimization is only done upon $F$. We use this model as an intermediary form to model our domain problems \cite{o2016toq,borle2019analyzing,o2018nonnegative,chang2019quantum}. A QUBO problem can alternatively be represented by an upper triangular matrix Q and a column vector $q$ of binary variables  with the objective to minimize the following expression:
\begin{align}
    F'(q) =  q^{T}Qq
\end{align}
Here, the qubit coefficient  $v_a$  from Eqn(\ref{eq:v}) is represented as a diagonal element $Q_{aa}$ and a quadratic coefficient $w_{ab}$ as an off diagonal elements $Q_{ab}$. For people curious about more on the subject, we recommend them the book by Crama and Hamer \cite{crama2011boolean}.

\subsubsection{Quantum computing}
 Quantum computing can be defined as leveraging quantum mechanical phenomena like superposition, entanglement, and quantum measurement for computational purposes \cite{nielsen2002quantum}. During the process of quantum computation, a qubit (quantum bit) can be 0 and 1 at the same time; with varying probabilities depending on how qubit is prepared.
 \begin{align}
     \ket{\psi} &= \alpha\ket{0} + \beta\ket{1}\label{eq:ket_psi}\\
     \text{where } \ket{0} &= \begin{pmatrix}
    1\\
    0
    \end{pmatrix} \text{and } \ket{1} = \begin{pmatrix}
    0\\
    1
    \end{pmatrix}\label{eq:ket_01}
 \end{align}

A qubit is described mathematically in Eqn(\ref{eq:ket_psi}). In a qubit, alpha and beta are amplitudes of a basis state (here 0 and 1) such that $Pr(0) = |\alpha|^{2}$, $Pr(1) = |\beta|^{2}$ and $|\alpha|^{2} + |\beta|^{2}=1$. The quantum state of  $n$ qubits can be represented by a column vector of length $2^n$.
 This quantum state vector is manipulated during its computation, commonly conceptualized using quantum gates, which take the mathematical expression of unitary matrices. The computation is designed in such a way that either the resultant quantum state is useful without measurement, or we can measure the desired bitstring from the quantum state with a high probability.
 
 The previous description is rudimentary in nature and is not comprehensive in fully articulating the process of quantum computing. But we hope it suffices for the scope of this paper.  We recommend the text by Nielsen and Chuang for a more detailed introduction of the subject \cite{nielsen2002quantum}. This work deals with quantum annealing, a subset of quantum computing focused on discrete optimization for the Ising model.
 
\subsubsection{Adiabatic quantum computing (AQC)}
This computational paradigm is based on the time dependent Hamiltonian evolution of a system. In Adiabatic Quantum Computing, the desired solution of our computation is expressed  as an eigenvector $\ket{\psi^{(P)}_{0}}$ of the Hamiltonian matrix $\hat{H}_P$ that corresponds to its smallest eigenvalue $\lambda^{(P)}_{0}$. This eigenvector would be called as a ground state of $\hat{H}_P$ and the eigenvalue $\lambda^{(P)}_{0}$ its energy.

The process begins with our quantum state as the ground state $\ket{\psi^{(B)}_{0}}$ of the initial Hamiltonian $\hat{H}_B$ whose ground state is easy to prepare as well as find analytically \cite{lucas2014ising} and evolves to the ground state of the problem solution $\ket{\psi^{(P)}_0}$. This is done based on the Adiabatic principle \cite{farhi2000quantum}, which ensures the quantum state remains in the ground state of the slowly varying $\hat{H}(t)$. The time dependant Hamiltonian $\hat{H}(t)$ becomes 
\begin{align}
\hat{H}(t) = (1 - t/T)\hat{H}_B + (t/T)\hat{H}_P \label{eq:aqc_pre}
\end{align}
Where $T$ is the maximum time allocated for the evolution to run. As $T \rightarrow \infty$, we get a state very close to the ground state of $\hat{H}_P$. To represent this better we can normalize the time variable as $s = t/T$ where $s \in [0,1]$.
\begin{align}
\hat{H}(s) = (1 - s)\hat{H}_B + s\hat{H}_P \label{eq:aqc}
\end{align}
One of the most important criterion for adiabatic quantum computing is the energy gap between the ground state energy and the first exicted state energy (the next consecutive energy in increasing order) of $\hat{H}(t)$. The duration of the adiabatic process $T$ and the minimum gap $g_{min}$ during the process are linked by the following relation
\begin{align}
    T &\gg \frac{\varepsilon}{g^{2}_{min}} \text{ or } T = O\left( \frac{1}{g^{2}_{min}} \right) \\
    \text{where } g_{min} &= \mintime_{s\in [0,1]} (E_{1}^{(s)} - E_{0}^{(s)})
\end{align}

Thus, a non-zero gap is crucial for us to obtain the ground state (best solution) of the problem Hamiltonian $\hat{H}_P$. Here, $\varepsilon$ is no larger than the maximum eigenvalue of the Hamiltonian $\hat{H}_P - \hat{H}_B$ and does not grow more than as a polynomial in the number of qubits \cite{farhi2000quantum}.

\subsubsection{Quantum annealing (QA)}
 Quantum annealing uses a quantum analog of simulated annealing that involves a transverse magnetic field to search the energy landscape of the problem Hamiltonian to find its ground state. For the Ising spin model, the variables $\sigma_a$ are mapped to the qubits spins.

The process begins with the qubits in an equal quantum superposition, meaning all potential qubit configurations have an equal probability of being measured. It then attempts to find the lowest energy configuration of the objective function $F(\sigma)$ by varying the transverse magnetic field strength (and gradually reducing it to 0). In the case of a physical quantum annealer, for long running-times, an ideal
finite-temperature annealer is would sample the Boltzmann distribution of the final Hamiltonian
at the annealer temperature($ \approx 1/\beta$) \cite{albash2017temperature}, where $\beta$ is the inverse temperature parameter, which governs the probability distribution of the spin configurations. This combination of superposition and time dependendent Hamiltonian evolution causes a tunneling effect between states. The Hamiltonian for D-wave's quantum annealing is as follows:
\begin{align}
\begin{split} \hat{H}(s) &= -\frac{A(s)}{2}\left( \sum_{a}\hat{\sigma}^{(x)}_{a} \right) \\
&+ \frac{B(s)}{2}\left(\sum_{a}h_{a}\hat{\sigma}^{(z)}_{a} + \sum_{a<b}J_{ab}\hat{\sigma}^{(z)}_{a}\hat{\sigma}^{(z)}_{b}\right)
\end{split}\label{eq:anneal_hamil}\\
\text{where } \hat{\sigma}_{a}^{(z)} &= (\otimes_{i=1}^{a-1}\hat{I}) \otimes (\hat{\sigma}^{(z)}) \otimes (\otimes_{i=a+1}^{n}\hat{I})\label{eq:sigmaz}\\
\hat{\sigma}_{a}^{(x)} &= (\otimes_{i=1}^{a-1}\hat{I}) \otimes (\hat{\sigma}^{(x)}) \otimes (\otimes_{i=a+1}^{n}\hat{I})\label{eq:sigmax}\\
\text{and }\hat{\sigma}^{(z)} &=
    \begin{pmatrix}
    1 & 0\\
    0 & -1
    \end{pmatrix} \text{ and } \hat{\sigma}^{(x)} =
    \begin{pmatrix}
    0 & 1\\
    1 & 0
    \end{pmatrix}\label{eq:pauli}
\end{align}
The $\hat{I}$ represents a $2\times 2$ identity matrix and $\otimes$ is the kronecker product operation. $A(s)$ and $B(s)$ are energies as a function of the normalized time $s$. If we put $A(s) = 2(s-1)$ and $B(s) = 2s$, you can see that Eqn(\ref{eq:aqc}) is a generalized form of Eqn(\ref{eq:anneal_hamil}). It also shows how the Ising cost function Eqn(\ref{eq:ising}) gets transformed to a Hamiltonian. We can Map the qubits in accordance with $\hat{\sigma}^{(z)}$ to scaler values of the Ising variables, we have $\ket{0} \rightarrow 1$ and $\ket{1} \rightarrow -1$. This is because $\hat{\sigma}^{(z)}$ has $-1$, $1$ as eigenvalues and $\ket{1}$, $\ket{0}$ as the respective eigenvectors.

Practically, it is not  clear if the D-wave quantum annealer adheres to the adiabatic principle completely \cite{shin2014quantum} (mostly due to the noise). Also, each machine has its own fixed anneal path. Since 2016, the main source of control over this path can be exercised by the pause, quench and reverse anneal features. Though these features are welcome and beneficial in a lot of cases \cite{chen2020and},  they do not allow for full control over the anneal procedure and are also subject to noise \cite{gardas2018defects}.
It is also worth noting that quantum annealing does not necessarily need an analog quantum processor such as the D-wave; we can discretize the annealing procedure for use on gate-model quantum computers as well \cite{willsch2020benchmarking}.
\subsubsection{Linear system of equations (LSE)}
A linear equation of $n$  variables $x \in \mathbb{R}^{n}$ is written in the following form:
\begin{align}
    a_{1}x_{1} + a_{2}x_{2} + ... + a_{n}x_{n} = b \label{eq:linear_eq}
\end{align}
A system of linear equations is a collection of  one or more equations involving the same variables, arranged as a matrix $A \in \mathbb{R}^{n\times n}$ a vector of variables $x \in \mathbb{R}^{n}$ and a vector $b \in \mathbb{R}^{n}$ such that
\begin{align}
    Ax = b \label{eq:lse}
\end{align}
The simplest method to solve LSE is by performing $A^{-1}b$. However, matrix inversion is seldom used for this task and LU decomposition  \cite{schwarzenberg1995matrix}, a type of Gaussian elimination  \cite{lay2018linear} is primarily used to solve for $x$ typically in $O(n^3)$ but as fast as $\approx O(n^{2.372})$ depending on the matrix multiplication algorithm used \cite{alman2021refined}.
\subsubsection{Linear least squares (LLS)}
Given a matrix $A \in \mathbb{R}^{m \times n}$, a column vector of variables $x \in \mathbb{R}^n$ and a column vector $b \in \mathbb{R}^m$ (Where $m>n$). The linear least squares problem is to find the $x$ that would minimize $\|Ax-b\|$ the most. In other words, it can be described as:
\begin{align}
    \argmin_x \|Ax - b\| \label{eq:LLS}
\end{align}
For a lot of practical cases, there exists no $x$ such that $Ax = b$. Various classical algorithms have been developed over the time in order to solve this problem. Some of the most prominent ones are (1)Normal Equations by Cholesky Factorization, (2) QR Factorization and the (3) SVD Method \cite{do2012numerically}.

\textbf{NOTE}: Although LSE and LLS is also applicable to complex numbers. The norm reduction based quantum annealing formulations are limited to only fixed point approximation of real values.
\subsubsection{Solving LSE and LLS with Quantum Annealing}
Both LSE and LLS can be viewed as norm minimization as represented by Eqn(\ref{eq:LLS}) (for LSE with unique or multiple solutions, $\|Ax - b\| = 0$ if $x$ is a solution). The following is the QUBO formulation for coefficients $v$ and $w$ when $x \in \{0,1\}^{n}$ \cite{o2016toq}.
\begin{align}
    v_{j} = \sum_{i}A_{ij}(A_{ij}-2b_{i})\label{eq:vbin}\\
    w_{jk} = 2\sum_{i}A_{ij}A_{ik}\label{eq:wbin}
\end{align}
The simplest way to approximate real values using fixed point approximation is by using a set of integer values $\Theta$ as follows \cite{borle2019analyzing},
\begin{align}
    x_{j} &\approx -2^{p+1} q_{j\emptyset} + \sum_{\theta \in \Theta} 2^{\theta} q_{j\theta}\label{eq:radix2_2comp}\\
    \text{where } \Theta &= \{l:l \in [o,p] \text{ where } l,o,p\in\mathbb{Z}\} \label{eq:Theta_set}
\end{align}
Here, $q_{j\emptyset}$ acts as a sign bit. The coefficients for this case are 
\begin{align}
    v_{js} &= \sum_{i} s A_{ij} (s A_{ij} - 2b_{i})\label{eq:v}\\
    w_{jskt} &= 2st \sum_{i}A_{ij}A_{ik} \label{eq:w}
\end{align}
where $s,t\in \vartheta \cup \{ 2^\theta : \forall \theta \in \Theta\}$. It should be noted that though more sophisticated methods to encode $x$ exist \cite{chang2019quantum,pollachini2021hybrid}, Eqn(\ref{eq:radix2_2comp}) for our purposes is sufficient to simulate and study the performance of quantum annealing for most of the other methods of encoding. The insights would apply in general to other works based on the principle of norm minimization.
\subsection{Related Work}
One of the seminal works in applying quantum computing to problems in linear algebra is the Harrow-Hassidim-Llyod (HHL) algorithm for solving LSE \cite{harrow2009quantum}. This gate-model algorithm has a exponential speedup over classical methods. However, there are various caveats that need to be mentioned \cite{aaronson2015read}: (i) the (normalized) vector $b$ needs to be efficiently prepared as a quantum state $\ket{b}$ (ii) matrix $A$ must be  sparse (iii) matrix $A$ must be well-conditioned (robustly invertible) and (iv) the vector $x$ is produced as a quantum state $\ket{x}$ and would need to undergo extensive tomography to extract all values (which would eliminate the exponential speedup). As a whole, research in this field is ongoing and extensive \cite{wiebe2012quantum,clader2013preconditioned,kerenidis2016quantum,childs2017quantum,wang2019quantum,kerenidis2020quantum,bravo2019variational,xu2021variational}.

As an alternative, formulations based on quantum annealing primarily focus on norm minimization such as in Eqn(\ref{eq:LLS}) by converting it into a QUBO formulation \cite{o2016toq,borle2019analyzing}. O'Malley et al.'s work was the first in using the D-wave for linear algebra problems, followed by other works such as for polynomial systems of equations \cite{chang2019quantum}, variants of non-negative matrix factorization \cite{o2018nonnegative,ottaviani2018low} and non-linear least squares \cite{chang2019least} among  others \cite{potok2021adiabatic,o2020tucker}. Quantum annealing offers a more efficient method for loading data onto a quantum computer and produces entirely classical results \cite{o2016toq,borle2019analyzing} that do not require a fault tolerant quantum computer. From an accuracy standpoint, iterative linear solvers based on quantum annealing perform comparably to classical solvers \cite{chang2019quantum,souza2022application}.

The main challenge with quantum annealing as a linear algebra solver is that it is a meta-heuristic, not an exact or even an approximate solver \cite{kadowaki1998quantum}. This is not a problem for linear algebra problems that are NP-Hard by themselves \cite{o2018nonnegative,ayanzadeh2020ensemble}. But for those that can be solved relatively efficiently by classical methods such as LSE and LLS, employing quantum annealing may be harder to justify. One aspect of this work is to state reasons and present supporting evidence based on hardware-agnostic simulations for specific cases where it may be viable to do so. These findings are useful for not only LSE and LLS problems, but also for a better understanding of problems that use a similar formulation \cite{o2018nonnegative,ottaviani2018low,chang2019least}.

\section{Adiabatic simulations for LLS problems} \label{sec:aqc_sim}
\subsection{Experiment setup}
As mentioned before, quantum annealing is a meta-heuristic that evolves the quantum state from a superposition of states and ends up as a classical state who's energy is defined by the classical Ising model like in Eqn(\ref{eq:ising}). Based on the time $T$ and the anneal schedule, it can potentially obey the adiabatic law and thus guarantee the ground-state (lowest energy) solution. Since the running-time is inversely proportional to the square of the minimum gap, by analyzing how the minimum gap shrinks, we can assess how the anneal time may increase.

In the work by Chang et al. \cite{chang2019least}, they presented a conjecture on how the gap would decrease when it comes to polynomial least squares, of which linear least squares is a subset. Among other things, their prediction was that quantum annealing time would scale linearly with the number of rows $m$ and exponentially with the number of precision bits $c$. These conjectures were made by analyzing how the ground and first excited state energy difference shrunk in the final Hamiltonian $\hat{H}_P$.

This however, is a very rough estimation of $g_{min}$ as it only depends on the energy difference in $\hat{H}_P$ and not any other factor (such as the choice of $\hat{H}_B$, the rest of the energies within $\hat{H}_P$ or even the anneal schedule). In order to get a better sense of how $g_{min}$ shrinks with the parameters, we conduct a set of simple theoretical simulations and see how the minimum energy gap changes according to the parameters of precision $c$, columns $n$, rows $m$ and the condition number $\kappa$. Our simulation  results show support for their prediction on the number of precision bits $c$ and contradicts their conjecture on the number of rows of data $m$. The details of these simulations are as follows:
\begin{itemize}
    \item \textbf{Gap $g_{min}$ as a function of no. of precision bits $c$}: We generate 100 problems of the type $A \in \mathbb{R}^{40 \times 2}$, $x \in \mathbb{R}^{2}$ and $b \in \mathbb{R}^{40}$. The solutions of these problems are integers in the range of $[-2,1)$, which can be represented with 2's-complement ( Eqn(\ref{eq:radix2_2comp})) and $\Theta = \{0\}$, i.e. $c = 2$. Although these problems can be solved sufficiently with 2 bits per entry in $x$, adding additional bits for precision should not affect our ability to represent the original ground state. We observe how $g_{min}$ behaves as we increase the bits of precision from $c=2$, $\Theta=\{0\}$  to $c=6$, $\Theta=\{4,3,2,1,0\}$.
    \item \textbf{Gap $g_{min}$ as a function of no. of variables $n$}: For number of variables $n \in \{2,3,4,5,6\}$, we generate 100 problems per $n$ of the type $A \in \mathbb{R}^{40 \times n}$, $x \in \mathbb{R}^{n}$ and $b \in \mathbb{R}^{40}$. Just like with the case above, these problems can be solved with a minimum of two bit precision with 2s complement and $\Theta = \{0\}$. We observe how $g_{min}$ behaves as we increase the number of variables $n$.
    \item \textbf{Gap $g_{min}$ as a function of number of rows $m$}: We generate 100 problems per $m \in \{10,20,30,...,290,300\}$ for $A \in \mathbb{R}^{m \times 4}$, $x \in \mathbb{R}^{4}$ and $b \in \mathbb{R}^{m}$. The minimum number of bits required for precision is $c=2$ similarly to how its done above. We observe how $g_{min}$ behaves as we increase the number of rows $m$.
    \item \textbf{Gap $g_{min}$ as a function of condition number $\kappa$}: We generate 100 problems per $\kappa \in \{1,2,3,...9,10 \} \cup \{20,30,40,...,290,300 \}$ for $A \in \mathbb{R}^{40 \times 4}$, $x \in \mathbb{R}^{4}$ and $b \in \mathbb{R}^{40}$ (for $c=2$ precision bits). We observe how $g_{min}$ behaves as we increase the condition number $\kappa$ of $A$ for problem Hamiltonians.
\end{itemize}

The problems are first converted into their QUBO form using Eqn(\ref{eq:v}) and Eqn(\ref{eq:w}) and then onto the Ising cost function by the substitution mentioned in Eqn(\ref{eq:ising2qubo}). After that all the $h$ and $J$ coefficients are scaled such that $h_a \in [-2,2]$ and $J_{ab} \in [-1,1]$. This is done since the D-wave machines usually have that range for their Ising coefficients. Even if the range of the coefficients were to change, it is safe to assume that quantum devices would be working with a finite range. For the annealing schedule, we assume $A(s) = 2(1-s)$ and $B(s) = 2s$ for Eqn(\ref{eq:anneal_hamil}) to make it completely compliant with Eqn(\ref{eq:aqc}). This is because the latter is a more standard form to assess adiabaticity in quantum computing \cite{farhi2000quantum}. We calculate the time-dependent Hamiltonian $\hat{H}(s)$ using Eqn(\ref{eq:anneal_hamil}) for 100 equally spaced discrete values in the interval $s \in [0,1]$. We can then calculate the energy difference $E^{(s)}_{1}-E^{(s)}_{0}$ for each discrete $s$ and approximate the $g_{min}$.

We also need to mention that all of these LLS problems have a solution $x$ such that $Ax = b$. This was done to circumvent the issue of limited precision for the current quantum optimizers. Although the  solutions to the generated problems are integers, this is not to be misinterpreted as finding integer solutions to linear systems, a different problem \cite{gomory2010outline}. The insights from these experiments would apply to LLS cases where $Ax \neq b$.

\subsection{Results and Discussion}
We describe our results visually with the help of Figures \ref{fig:gmin_prec} to \ref{fig:gmin_cond2}. In these plots, a median of the  minimum gap $g_{min}$ is plotted and the error bars represent the median absolute deviation (MAD) observed when calculating the $g_{min}$.
\subsubsection{Running-time as a function of no. of precision bits $c$ of $x$}

\begin{figure}[h]
    \centering
    \includegraphics[width=7.31cm,height=5.5cm]{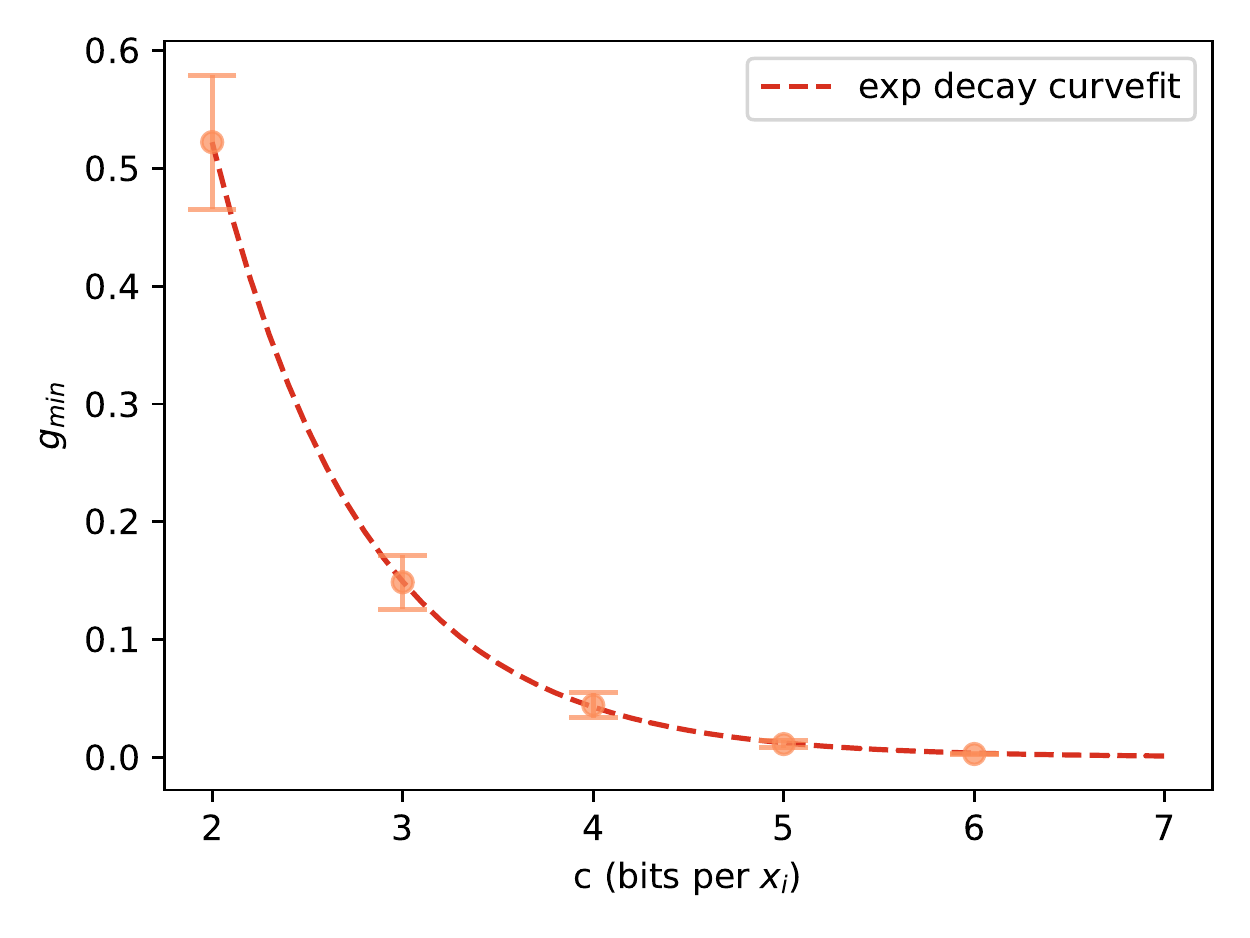}
    \caption{Plotting $g_{min}$ as a function of bits of precision $c$. The size of the Hamiltonians ranges from 4  qubits (for $c=2$) to 12 qubits (for $c=6$). The exponential curve type $ae^{-bx}$ fits best with a relative error of $\sim0.003$ with $a \approx 6.407$ and $b \approx 1.2534$. }
    \label{fig:gmin_prec}
\end{figure}
Figure \ref{fig:gmin_prec} plots $g_{min}$ for the number of precision bits used in the QUBO formulation. Due to the size of the Hamiltonian matrix growing exponentially, we could only conduct the experiment for 5 values of $c$ (100 test cases for each).
Across all our simulations, the strongest relation appears to be that $g_{min}$ drops exponentially as precision is increased. Thus, for future adiabatic computing approaches and depending upon the implementation, there would be a point where increasing the precision would hinder the effort to find the ground state (by making the running-time exponentially long).
It does not mean that quantum annealing can't be used for LLS problems that require high precision, since there are iterative techniques that allow for arbitrary precision \cite{chang2019quantum} by redefining the interval covered by a set of qubits.  But for a given anneal run, the number of precision bits participating in the process will be low.

\subsubsection{Running-time as a function of no. of variables $n$ in $x$}
\begin{figure}
    \centering
    \includegraphics[width=7.31cm,height=5.5cm]{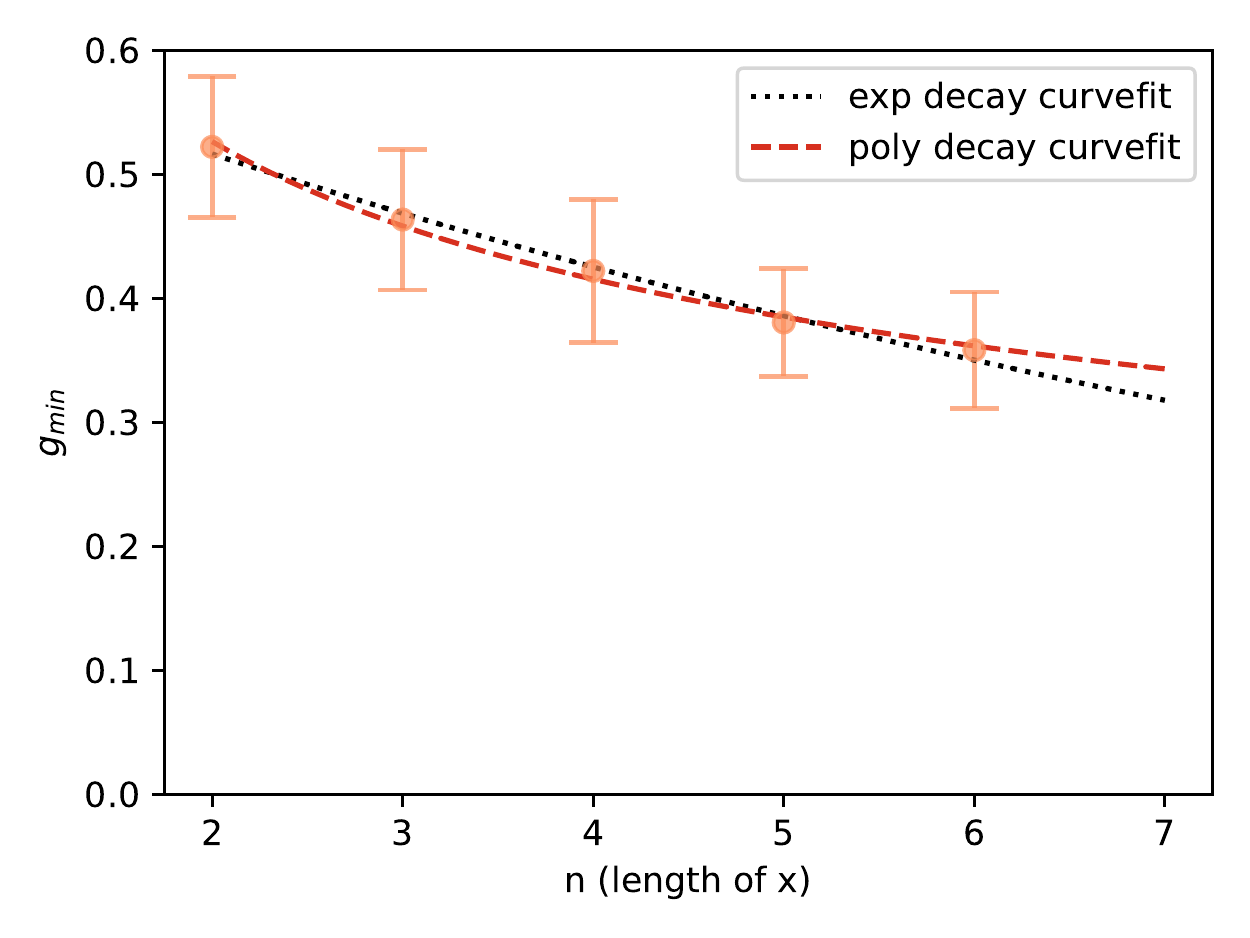}
    \caption{Plotting $g_{min}$ as a function of variables $n$. The size of the Hamiltonians ranges from 4  qubits (for $n=2$) to 12 qubits (for $n=6$). We fit both exponential ($ae^{-bx}$, $a\approx0.6273$ and $b\approx0.0970$) and polynomial ($ax^{-b}$, $a\approx0.6673$ and $b\approx0.3415$) decay curves with relative errors of $\sim0.013$ and $\sim0.011$ respectively.}
    \label{fig:gmin_n}
\end{figure}

If we fix the precision and plot a decay curve for the relation of $g_{min}$ with the number of variables $n$, a polynomial curvefit gives a minutely better relative error than an exponential one (see Figure \ref{fig:gmin_n}). However, because we could only test on a small $n$  due to the exponential size of the matrices to simulate, it is hard to predict. On a conservative estimate, it is possible that the relation is exponential. Since an integer programming formulation can also be extended from Eqn(\ref{eq:v}) and Eqn(\ref{eq:w}). But it should also be noted that the exponential decay curve fitted here has a far gentler slope than the one for precision.

\subsubsection{Running-time as a function of no. of rows $m$ in $A$ }
\begin{figure}
    \centering
    \includegraphics[width=7.31cm,height=5.5cm]{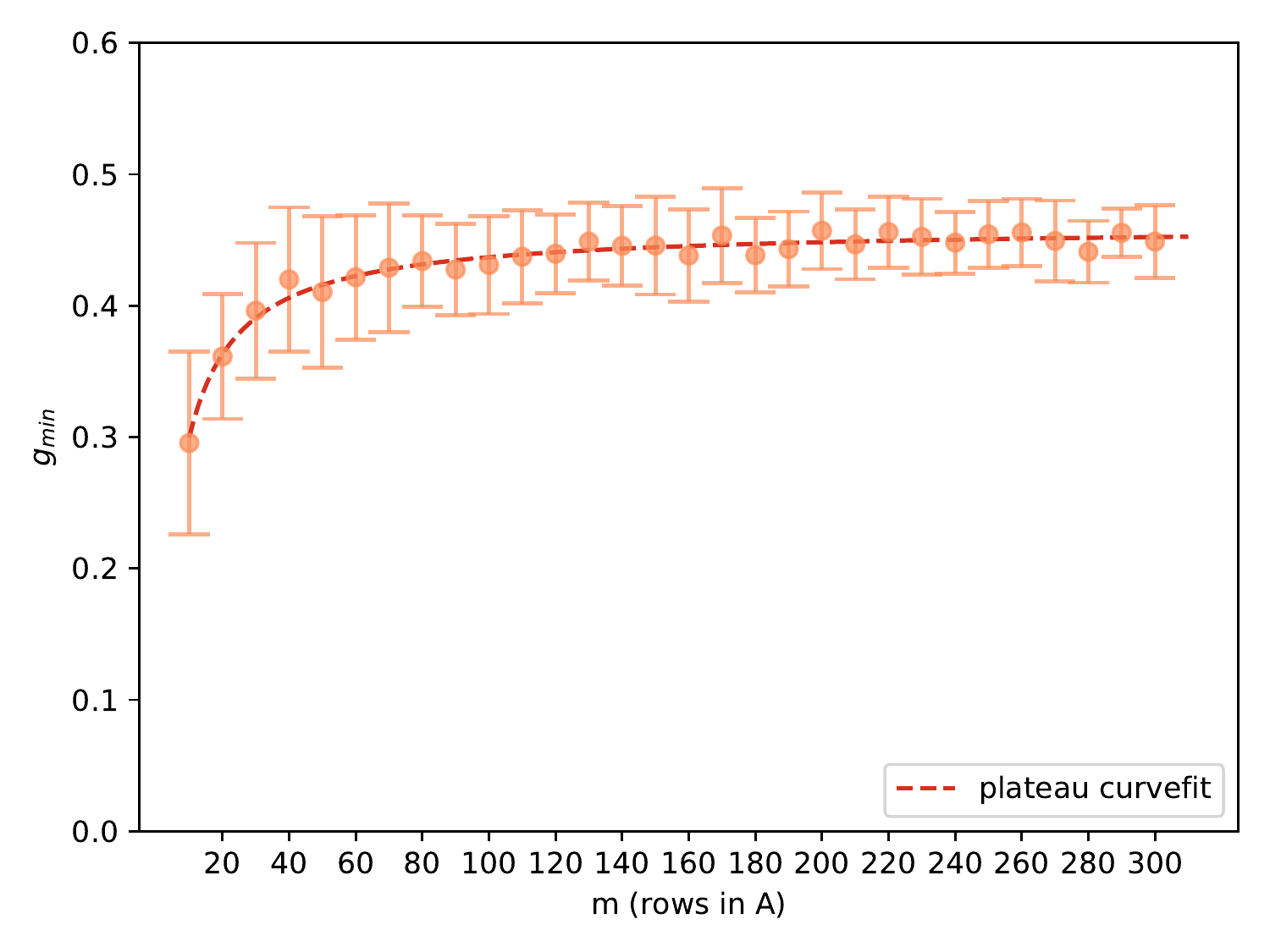}
    \caption{Plotting $g_{min}$ as a function of variables $m$. The size of the Hamiltonians is 8 qubits for all the datapoints (for $n=4$,$c=2$).  We fit a plateau curve ($ax/(b+x)$, $a\approx0.4603$ and $b\approx5.355$) with a relative error of $\sim0.013$.}
    \label{fig:gmin_m}
\end{figure}

The results from attempting to plot $g_{min}$ as a function of rows $m$ shows a gradual increase in $g_{min}$ and a potential plateau. The explanation for this could be the nature of the dataset. Our matrices were dense and though they had positive as well as negative entries, calculating the QUBO coefficients $v$s and $w$s using Eqn(\ref{eq:v}) and Eqn(\ref{eq:w}) can increase $g_{min}$ as the number of rows increases (since there is more data to add up).

While, we can't claim that the relation between $m$ and $g_{min}$ is one in which anneal time decreases as  the number of rows increases, it is likely that the number of rows don't play a direct role in the size of $g_{min}$ . If we consider the case of sparse matrices, there would be very few non-zero entries in each column of $A$. Although matrix sparsity may affect how fast $g_{min}$ grows, it seems unlikely it would start shrinking $g_{min}$. It may be more likely that $g_{min}$ does not get affected by $m$ in the general case.

We say this because of two reasons: (i) The size of the Hamiltonian itself does not depend on $m$ and (ii) Because  Eqn(\ref{eq:v}) and Eqn(\ref{eq:w}) show no direct relation for an increase in $m$ reducing the coefficients (and $g_{min}$ consequently). It may turn out that there would exist a subset of the problem for which $g_{min}$ has a more clear relation to the number of rows being processed. But the evidence doesn't seem to suggest so in the general case (as also in  the work done on the D-wave machine in \cite{potok2021adiabatic}).

\subsubsection{Running-time as a function of condition number $\kappa$ of $A$}
\begin{figure}[h]
    \centering
    \includegraphics[width=7.31cm,height=5.5cm]{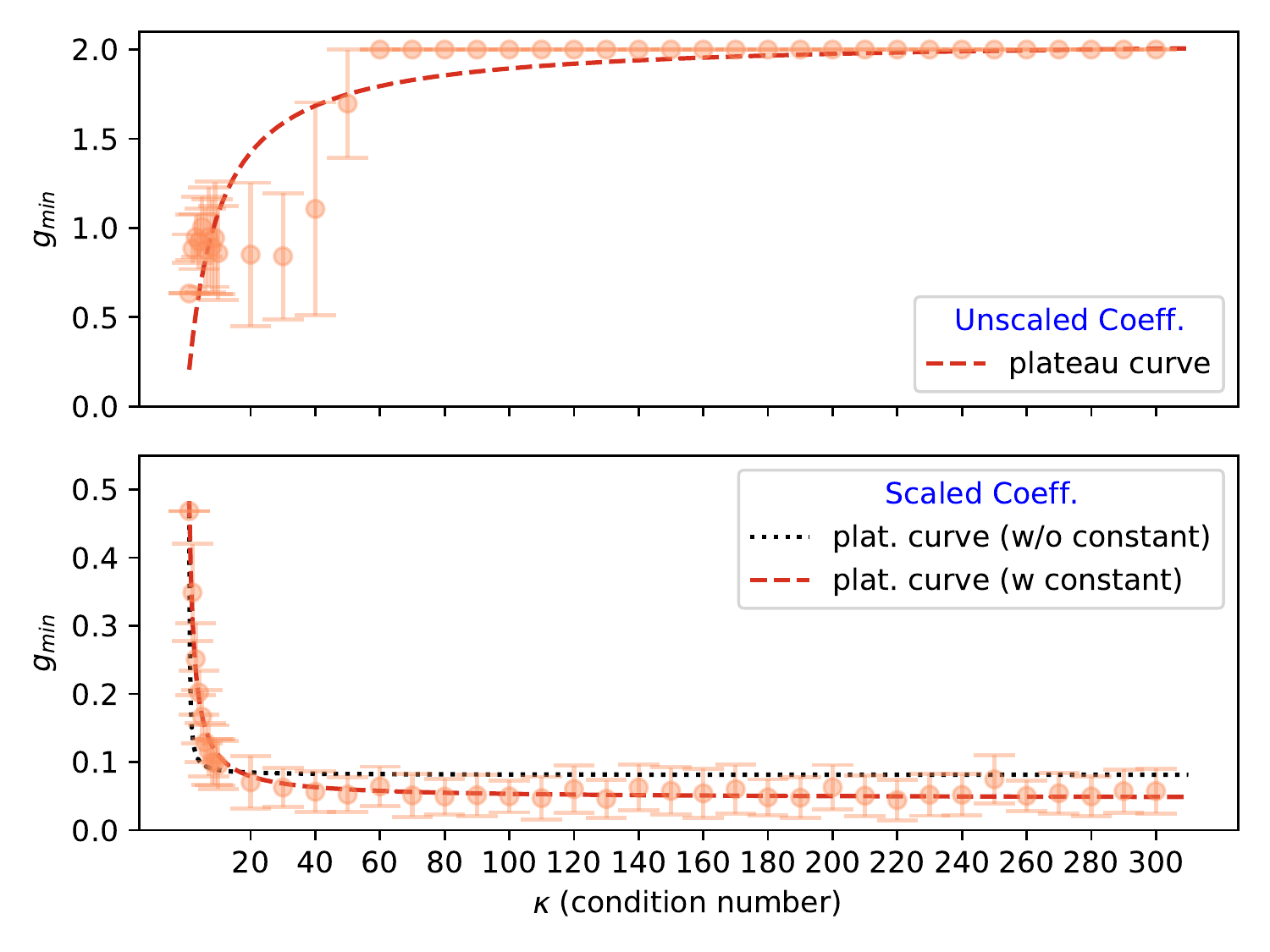}
    \caption{Plotting $g_{min}$ as a function of condition number $\kappa$. (TOP) is when the Ising coefficients of the Hamiltonians are not scaled and (BOTTOM) is where the Ising Hamiltonians are scaled such that $h\in[-2,2]$ and $J\in[-1,1]$. The size of the Hamiltonians is 8 qubits for all the datapoints (for $n=4$,$c=2$).  We fit plateau curves  without and with an offset of types $ax/(b+x)$ and $ax(b+x) + c$ respectively.}
    \label{fig:gmin_cond1}
\end{figure}

Finally, just like with the effect of rows $m$ on the running-time, the condition number $\kappa$ of the matrix $A$ is another parameter that doesn't affect the size of the Hamiltonian, but can impact the performance of the process. In a lot of algorithms, both quantum and classical \cite{harrow2009quantum,do2012numerically,van1992bi,wiebe2012quantum}, the conditioning of the matrix can be a real impediment in solving the linear system or linear least squares problem. In contrast, our simulations show that the Ising model based adiabatic evolution seems to be mostly resilient to ill-conditioned systems. This can be seen in the Figure \ref{fig:gmin_cond1} in which the minimum energy gap $g_{min}$ eventually plateaus in both the subplots. We have two subplots here because in addition to the standard scaled Ising coefficients, we also compare with the (as is)  unscaled problems. This is done in order to closely examine the results of the scaled problems, in which the $g_{min}$ sharply drops till $\kappa = 10$ only to gradually plateau by around $\kappa=70$. If you look at the unscaled subplot, it appears like the mirror opposite of the scaled one, which peaks at $g_{min}=2$, the energy gap between the ground and first excited states of the starting Hamiltonian $\hat{H}_B$. This means that at a condition number $\kappa$, the unscaled adiabatic quantum process is not limited by the energy gap in the final Hamiltonian $\hat{H}_{P}$.

\begin{figure}
    \centering
    \includegraphics[width=7.31cm,height=5.5cm]{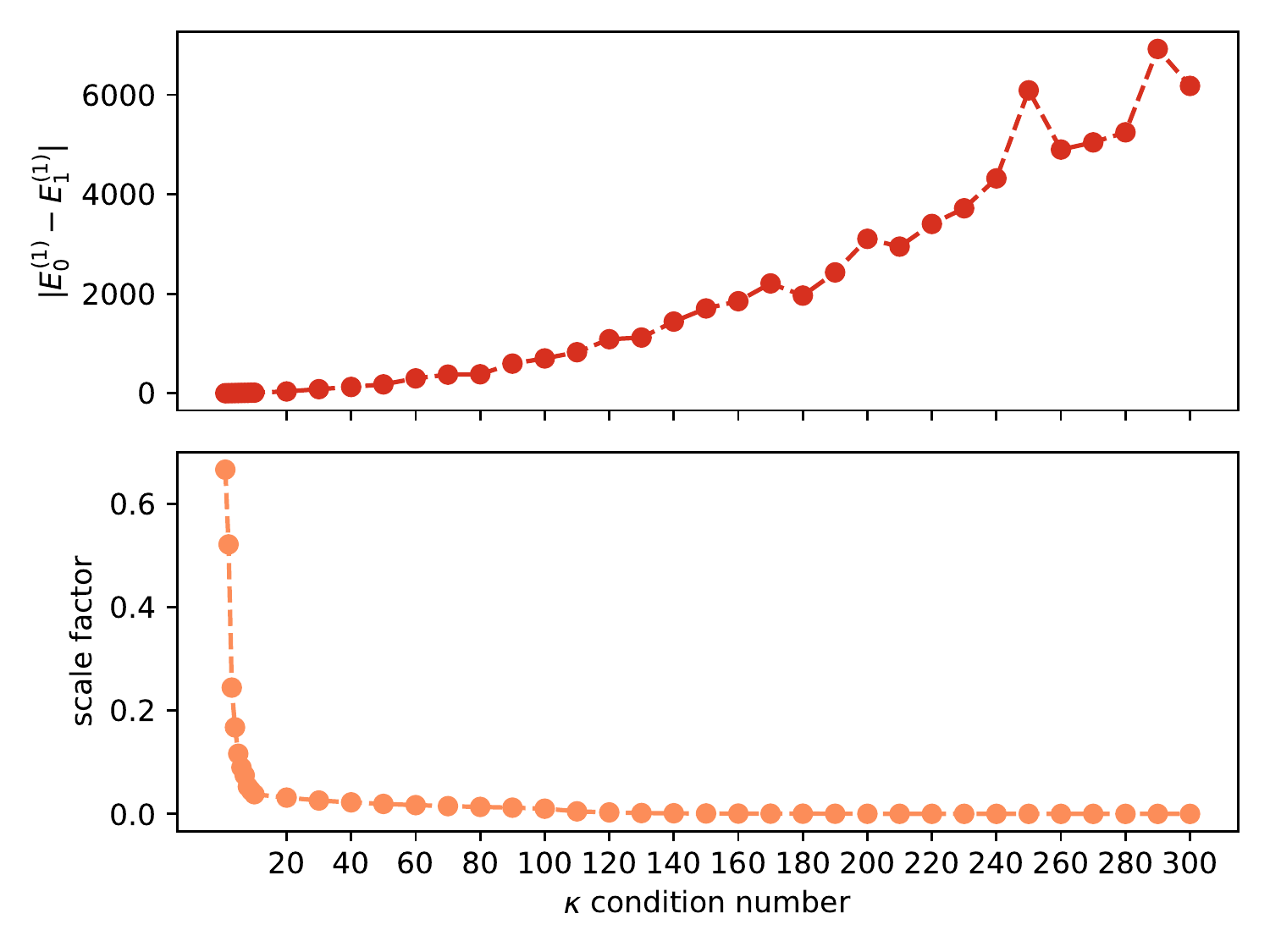}
    \caption{(TOP) Plotting the unscaled energy difference between the ground state and the first excited state of the problem Hamiltonian $\hat{H}_P$ and (BOTTOM) plotting the scale factor for scaling the Ising coefficients in a range of $h\in[-2,2]$ and $J\in[-1,1]$ for those Hamiltonians, against the condition number $\kappa$.}
    \label{fig:gmin_cond2}
\end{figure}

In order to study the phenomenon in Figure \ref{fig:gmin_cond1}, we conduct a corollary simulation in which we plot the unscaled energy gap of the problems at the final Hamiltonian $\hat{H}_P$  given by $| E_{0}^{(1)} - E_{1}^{(1)}|$ and the scale factors that would be required for scaling the problems down to the above mentioned range as a function of the condition number $\kappa$ of the matrix $A$ (all values are medians). You can see in Figure \ref{fig:gmin_cond2} that while the energy gap keeps growing at an almost predictable rate, the scale factor plummets between $\kappa=1$ to $\kappa=10$, only to then reduce at a much slower rate afterwards.

In terms of applying this practically, it would make sense that most physical annealers would require our Ising problems to be scaled down to a particular range. And thus, the running-time would expect to increase sharply till a particular condition number $\kappa$ after which the scaling should improve significantly. This observation is corroborated in the work by Chang et al. who iteratively solved linear systems on a D-wave annealer \cite{chang2019quantum} for a much smaller range of $\kappa$.  However, other practical issues would also affect the performance; such as round-off errors during the  QUBO/Ising preparation and the hardware precision for the Ising coefficients.

\section{Quantum-Assisted Linear Solver}\label{sec:qals}
The previous section focused on numerical simulations for assessing how might the running-time for quantum annealing scale with relation to different parameters. In this section, we want to propose a hybrid quantum-classical approach of using quantum annealing for an approximate solution $x^{(q)}$. This approximate solution will be the initial guess $x^{(c)}_{0}$ for a classical fixed point iteration method (such as bi-conjugate gradient) for convergence to an acceptable error threshold. The fundamental concept of this work was inspired by the work on quantum-assisted greedy algorithms \cite{ayanzadeh2019quantum}.

\begin{algorithm}
\caption{Quantum-assisted linear solver}\label{alg:qals}
\begin{algorithmic}[1]
\Procedure{MAIN}{$A,b,\Theta,f(x)$}
\State Initialize QUBO matrix $Q \gets \emptyset$
\State Use Eqn(\ref{eq:v}) and Eqn(\ref{eq:w}) with $A,b$ and $\Theta$ to prepare the QUBO matrix $Q$
\State Use a Quantum Annealer QA with $Q$ to produce $x^{(q)} \gets \text{QA}(Q)$ \label{line:qa}
\State Use the result from QA  as the initial guess for the classical solver $x^{(c)}_{0} \gets x^{(q)}$ 
\State $i \gets 1$
\While{termination criteria not reached} 
    \State Use $f(x)$ to produce new result, $x^{(c)}_{i}  \gets f(x^{(c)}_{i-1})$
    \State $i \gets i + 1$
\EndWhile
\State  \textbf{return} $x^{(c)}_{i}$
\EndProcedure
\end{algorithmic}
\end{algorithm}

Algorithm \ref{alg:qals} shows a general algorithmic description of this hybrid approach. Here, $f(x)$ refers to a fixed point iteration method like LSMR for LLS or BiCG for LSE. This general approach can also be extended and developed for other problems such as non-negative matrix Factorization \cite{gillis2017introduction}. Because a quantum annealer's own precision is often limited, we can also choose to iteratively use QA in line \ref{line:qa} with adaptive coefficients like in \cite{chang2019quantum} until a reasonably good quantum based result is produced.
\subsection{Experiment setup}
We randomly generate a dataset of 50 LSE and 50 LLS problems for sizes $n\in \{16,35\}$ (a total of 200 problems). The number of rows $m$ for LLS is 100 ($m=n$ for linear systems). Here, $x$ is such that $Ax = b$ for both types of problems and all entries in x are integers in the range of $[-8,8)$, which makes $\Theta = \{2,1,0\}, c=4$. This is for the same reason as mentioned in the previous section. And like before, we believe the core learning will also apply to the $Ax \neq b$ LLS problems.

We use the D-wave quantum annealers: 2000Q and the Advantage 4.1 Performance Upgrade (APU) for our experiments. The 2000Q, an older system, is limited to only problems where $n=16$ for our precision requirements. With an anneal time of $5 \mu s$ and a total of 1000 runs for each problem, we generate our approximate quantum results $x^{(q)}$ use it as the initial guess $x^{(c)}_{0}$ for the classical fixed point iteration methods.

For the classical part, our method of choice is the least squares MINRES (LSMR) method for LLS \cite{fong2011lsmr} and the bi-conjugate gradient (BiCG) method for LSE \cite{fletcher1976conjugate}. We use the SciPy implementation of these methods with the default termination criteria \cite{2020SciPy-NMeth} and compare against an initial guess of an all zero vector $x^{(c)}_0 \gets \{0\}^{n}$. Since the size of these problems that can fit on these devices is still relatively small, we cannot observe any practical running-time advantage as of the time of writing. Our metrics for comparison are (i) the number of iterations to converge to the termination criteria and (ii) the absolute error $\|Ax-b\|$ in the final result for a common number of iterations. For the second metric of comparison, we choose the number of iterations required to converge when $x^{(c)}_{0} \gets \{0\}^{n}$  as the common number of iterations for both the approaches. 

\subsection{Results and Discussion}
\begin{figure}
  \centering
  \includegraphics[width=7.31cm,height=5.5cm]{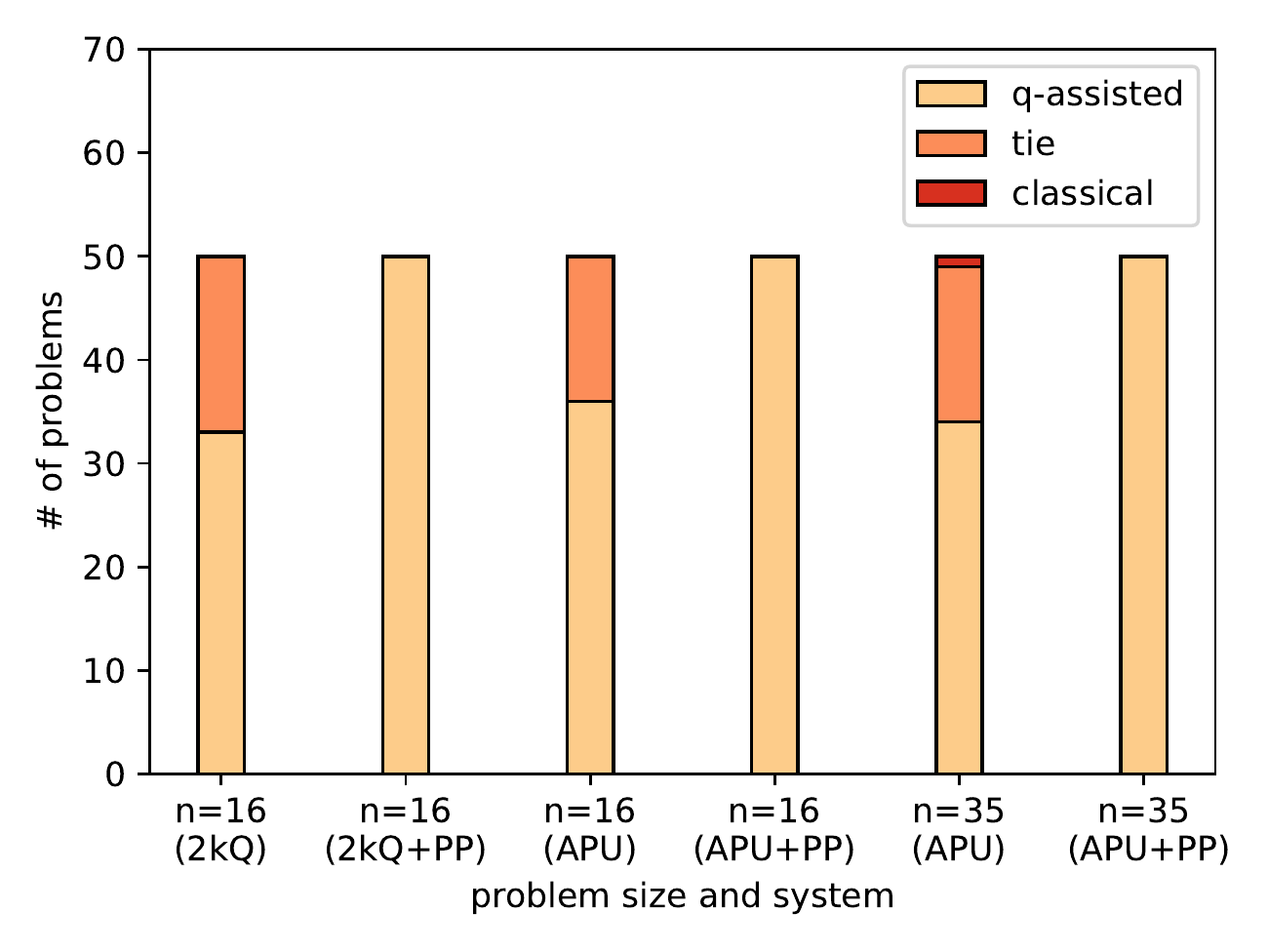}  
  \includegraphics[width=7.31cm,height=5.5cm]{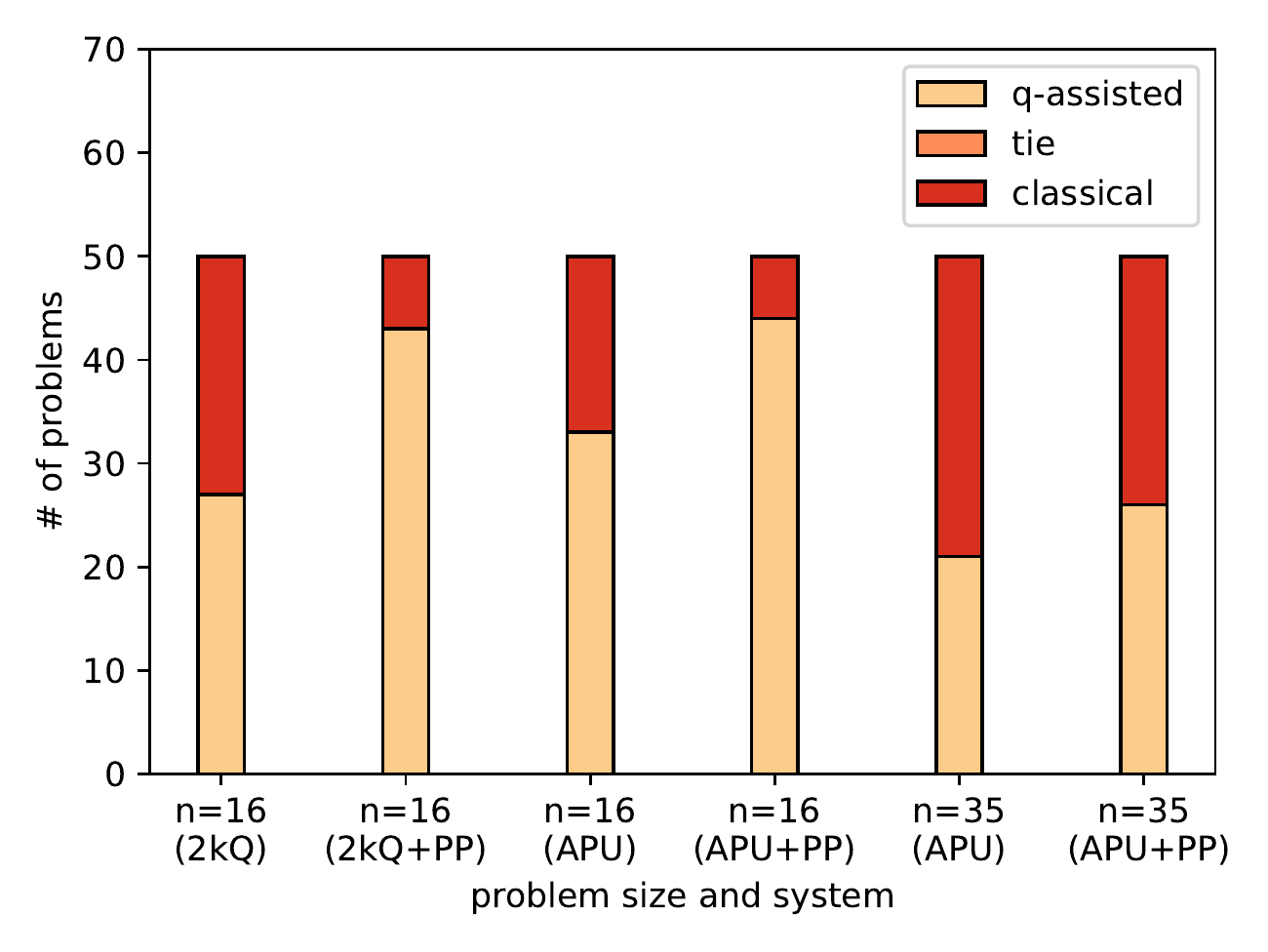}  
\caption{Bar graph for showing the count of LLS problems for which either quantum-assisted ($x^{c}_{0} \gets x^{(q)}$)  or classical ($x^{c}_{0} \gets \{0\}^{n}$) initial guess had an advantage, or if it resulted in a tie. (TOP) is when comparing for the number of iterations and (BOTTOM) is when comparing for absolute error with fixed iterations (less is better in both cases). 2kQ refers to D-wave 2000Q and APU refers to D-wave Advantage 4.1 (aka Advantage Performance Update). PP indicates the greedy-descent post-processing technique applied on quantum results}
\label{fig:lsmr_res}
\end{figure}

\begin{figure}
  \centering
  \includegraphics[width=7.31cm,height=5.5cm]{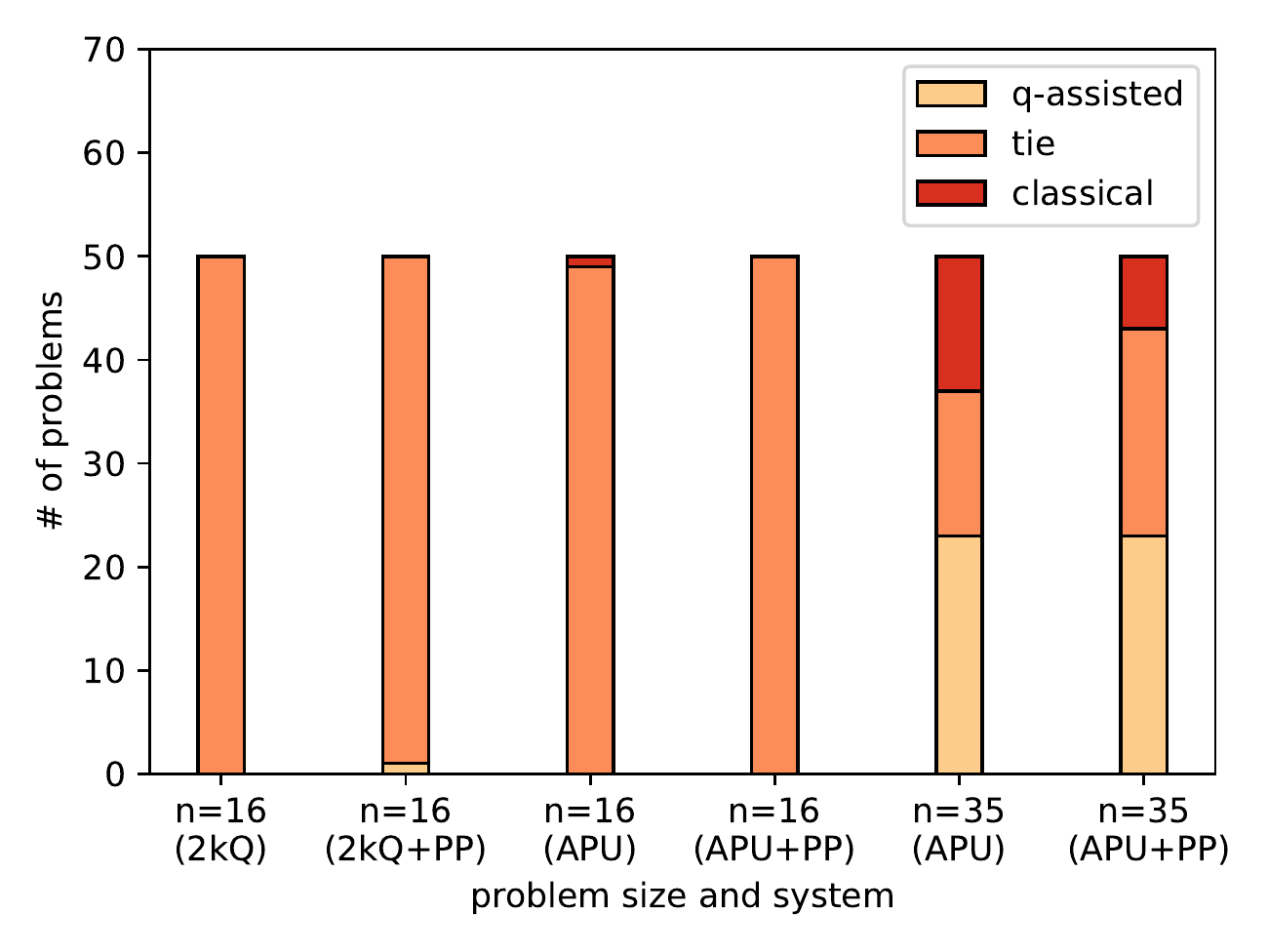}  
  \includegraphics[width=7.31cm,height=5.5cm]{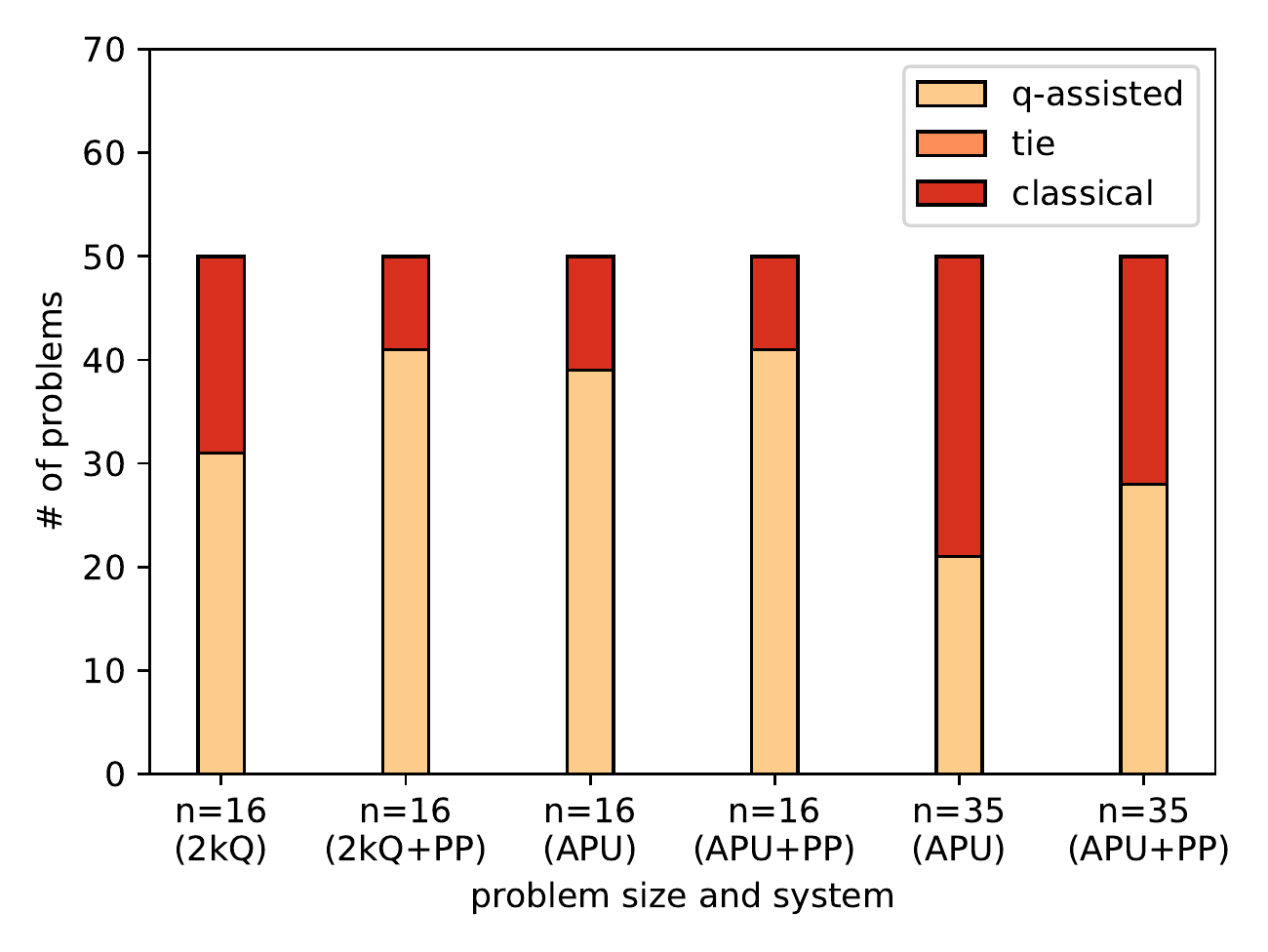}  
\caption{Bar graph for showing the count of LSE problems for which either quantum-assisted ($x^{c}_{0} \gets x^{(q)}$) or classical ($x^{c}_{0} \gets \{0\}^{n}$) initial guess had an advantage, or if it resulted in a tie. (TOP) is  when comparing for the number of iterations and (BOTTOM) is when comparing for absolute error with fixed iterations (less is better in both cases). 2kQ refers to D-wave 2000Q and APU refers to D-wave Advantage 4.1 (aka Advantage Performance Update).  PP indicates the greedy-descent post-processing technique applied on quantum results}
\label{fig:bicg_res}
\end{figure}

Figure \ref{fig:lsmr_res} and \ref{fig:bicg_res} show the results of our experiments for LLS and LSE problems respectively. For the quantum-assisted results, We also used the standard greedy descent post-processing technique (provided by D-wave's API called {\tt dwave-greedy}) on the results from the D-wave quantum annealers, and show results for both the raw and post-processed results.

From our results, we see that the quantum-assisted method is more effective for LLS problems than LSE problems. We conjecture the reason for that to be two folds: (i) based on the number of rows impacting the $g_{min}$ (refer Figure \ref{fig:gmin_m}) and (ii) a difference in how the termination criteria is evaluated in those two different problem types. We will get a better idea as the system size for quantum optimizers increases. For LSE however, we only saw an advantage for the quantum-assisted approach (in terms of iterations) for $n=35$. However, for the fixed number of iterations case, LSE was effective even for $n=16$ on the 2000Q machine.

A counter point to this however, is that for the cases where quantum-assisted guess was better, only a $\approx9\%$ improvement in iterations was observed for LLS where $n=16$. This is a small improvement, perhaps due to our small anneal time of $5 \mu s$ (the default anneal time is $20 \mu s$ for D-wave annealers). This reduces to $\approx5\%$ for $n=35$. So though a similar number of problems benefited from the quantum-assisted guess, the improvement margin declined. It may  be natural to expect a drop-off in improvement if the anneal time is being kept constant and the size of the problem is being increased. Further research would be required on future systems (that are larger) to assess this trend\footnote{all the percentage improvements mentioned were for raw results without post-processing}.  A median improvement of $\approx8\%$ was observed for $n=35$ LSE problems. 

In terms of comparing the 2000Q and the APU machines themselves, the 2000Q machine does not perform as well as the APU for $n=16$. This is probably because the on-chip qubit-connectivity of the Advantage machine is higher than the 2000Q. In essence, when mapping the same QUBO to the actual chip via minor-embedding \cite{choi2008minor}, it results in more number of qubits being used for the 2000Q than in the Advantage machine, which increases the size of the problem.

The greedy descent post-processing method was also more effective  for the LLS problems. In some cases, we obtained the exact solution without using LSMR. While our results are promising, we have to keep in mind that the qubo preparation for this problems has a cubic complexity: $O(mn^2)$ for LLS, $O(n^3)$ for LSE. Since LSE can be solved in as fast as $\approx O(n^{2.372})$, it is yet another indication why LSE may not be the best use-case problem for quantum annealing.

For LLS problems. In a previous work \cite{borle2019analyzing}, it was shown how quantum annealing for LLS can be faster than certain direct methods for LLS (under certain assumptions). However, when comparing quantum annealing assisted solvers against purely classical iterative methods, it may depend on various other factors like the termination criteria, the iteration method chosen, the distance from the initial guess and the final solution and many more. We hope our contribution will help the community for future research, particularly as the qubit capacity of quantum annealers increase.

\section{Conclusion}\label{sec:conc}
In our work, we attempt to address the viability of using the quantum annealing meta-heuristic for linear least squares and linear systems of equations. While previous works have mostly focused on empirical work on the machines, we here simulate the adiabatic quantum evolution to observe the effect of different parameters of the above problems on the running-time. These simulations are important since the phenomena observed are not vendor specific and are based on the foundational principles of quantum annealing and adiabatic quantum computing. Our simulations show that quantum annealing is promising for ill-conditioned systems and linear least squares problems when rows far exceed the number of columns $m\gg n$, our simulations also show that the running-time has an exponential slowdown as the number of bits of precision for variables are increased. These results show corroboration of evidence which while having being previously observed, wasn't conclusive by itself. It still would take more work to come to any definite conclusions, but we hope our work is a step in that direction. 

In our other contribution, we propose a hybrid quantum-classical method for solving linear systems of equations and linear least squares. Here, a quantum annealer was used to generate an approximate solution in a short amount of time and then a classical fixed point iteration method helped to solve the problem to convergence. Our results show  this approach being effective more for the least squares problems than the linear systems, though further work would be needed. We hope this general approach may also be useful for other harder problems that may be more suitable for quantum annealing.
\bibliographystyle{IEEEtran}
\bibliography{references}

\end{document}